\documentclass[runningheads]{llncs}
\usepackage{graphicx}
\usepackage{marvosym}
\usepackage{multirow}
\usepackage{tabu}
\usepackage{bbm}
\usepackage{diagbox}
\usepackage[english]{babel}
\usepackage{comment}
\usepackage{array}
\usepackage{amstext}
\usepackage{makecell}
\usepackage{tablefootnote}
\usepackage{babel}
\usepackage{booktabs} 
\usepackage{color}
\usepackage[table]{xcolor}
\usepackage{amsmath}
\usepackage[caption=false]{subfig}
\usepackage{mathrsfs}
\usepackage{amsfonts}
\newcommand{\ie}{\textit{i.e.}}

\begin{document}
\title{Scale-aware Test-time Click Adaptation for Pulmonary Nodule and Mass Segmentation}
\titlerunning{Test-time Click Adaptation for Pulmonary Lesion Segmentation}

\author{Zhihao Li\inst{1, 3, 4, 5}$^*$ \and
        Jiancheng Yang\inst{2, 6}$^*$ \and
        Yongchao Xu\inst{1, 3}$^*$$^*$ \and
        Li Zhang\inst{6} \and
        Wenhui Dong\inst{1, 3} \and
        Bo Du\inst{1, 3, 4, 5}$^*$$^*$}
\authorrunning{Z. Li et al.}
%
\institute{School of Computer Science, Wuhan University, Hubei, China \\ 
        \email{(dubo@whu.edu.cn, yongchao.xu@whu.edu.cn)} \and
        Computer Vision Laboratory, Swiss Federal Institute of Technology Lausanne (EPFL), Lausanne, Switzerland \and  
        Artificial Intelligence Institute of Wuhan University, Hubei, China\and
        Hubei Key Laboratory of Multimedia and Network Communication Engineering, Hubei, China \and
        National Englineering Research Center for Multimedia Software, Hubei, China \and
        Dianei Technology, Shanghai, China}

\renewcommand{\thefootnote}{\fnsymbol{footnote}} 
\footnotetext{\hspace*{2mm}$^*$ Equal contributions: Zhihao Li and Jiancheng Yang.\\
                $^*$$^*$ Corresponding authors contributed equally: Bo Du (dubo@whu.edu.cn) and Yongchao Xu (yongchao.xu@whu.edu.cn).} 
\renewcommand{\thefootnote}{\arabic{footnote}}
\maketitle              
\begin{abstract}

Pulmonary nodules and masses are crucial imaging features in lung cancer screening that require careful management in clinical diagnosis. Despite the success of deep learning-based medical image segmentation, the robust performance on various sizes of lesions of nodule and mass is still challenging. In this paper, we propose a multi-scale neural network with scale-aware test-time adaptation to address this challenge. Specifically, we introduce an adaptive \textit{Scale-aware Test-time Click Adaptation} method based on effortlessly obtainable lesion clicks as test-time cues to enhance segmentation performance, particularly for large lesions. The proposed method can be seamlessly integrated into existing networks. Extensive experiments on both open-source and in-house datasets consistently demonstrate the effectiveness of the proposed method over some CNN and Transformer-based segmentation methods. Our code is available at \url{https://github.com/SplinterLi/SaTTCA}.

\end{abstract}
%
%
%

\section{Introduction}

\begin{figure}[tb]
	\includegraphics[width=\linewidth]{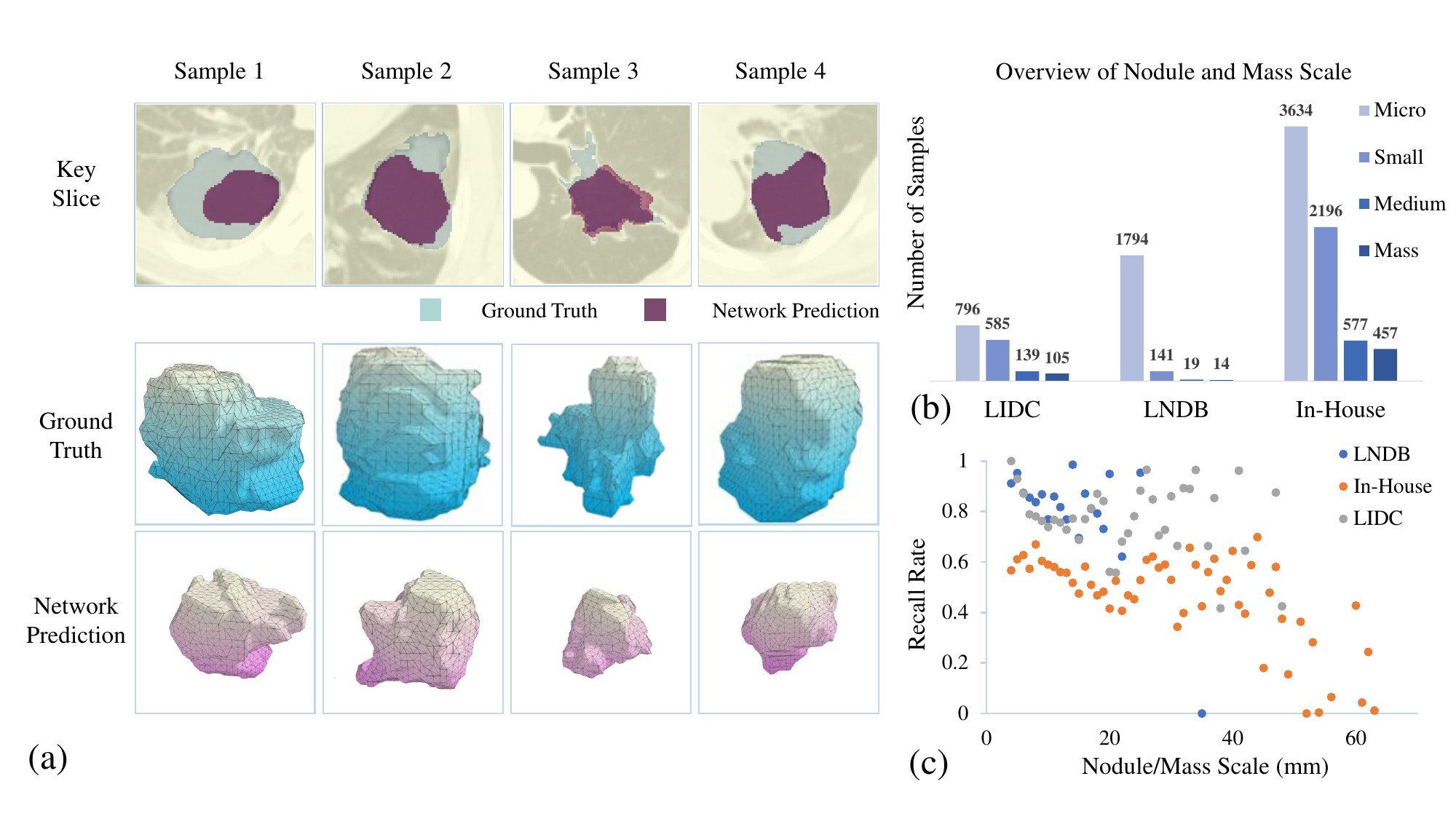}
	\caption{\textbf{(a)}: Visualization on results of four large-scale mass segmentation given by nnU-Net baseline~\cite{isensee2021nnu}. Compared with the ground-truth segmentation, the recall rate for these four samples is 46.29\%, 58.34\%, 79.51\%, and 68.51\%, respectively. This is significantly lower than the mean value of 81.68\%. \textbf{(b)}: Statistics of the number of nodules at different scales in three datasets. The range of nodule diameter corresponding to Micro, Small, Medium, and Mass is (0, 10], (10, 20], (20, 30], [30, $\infty$), respectively. \textbf{(c)}: The distribution of recall rate with respect to the nodule size. Existing methods have low recall rates for the segmentation of large scale nodules and masses.}
	\label{fig:introduction}
\end{figure}

Lung cancer is the main cause of cancer death worldwide~\cite{sung2021global}. Pulmonary nodules and masses are both features present in computed tomography images that aid in the diagnosis of lung cancer. The primary difference is that a nodule is smaller than 30 mm in diameter, while a mass is larger than 30 mm~\cite{vachani2022probability}. Early detection of these features is crucial to aid physicians in making a diagnosis of benign or malignant tumors~\cite{yang2020hierarchical} and determining follow-up treatment. Lesion segmentation can be utilized to evaluate two important factors: the volume of the lesion and its growth rate~\cite{Heuvelmans2013OptimisationOV,gould2013evaluation,macmahon2017guidelines,li2020learning}. Furthermore, obtaining accurate information regarding the nodule can assist in determining the appropriate resection method and surgical margin required to preserve as much lung function as possible.~\cite{schuchert2007anatomic,oizumi2011anatomic}.

Segmenting nodules is a tedious task that requires significant human labor. Computer aided diagnosis (CAD) systems can significantly reduce such heavy workloads. The accuracy of the existing nodule detection model reaches 96.1\%~\cite{liu2022stbi} accuracy. However, the accuracy of the 3D nodule segmentation model is prone to significantly decline in the application, regardless of whether its structure is based on CNN or Transformer~\cite{azad2022transdeeplab}. As shown in Fig.~\ref{fig:introduction} (a-c), the recall rate of the large-scale nodule and mass is usually lower than the average level. The main reason is that the lesion scale in the two public datasets are relatively small, which matches the fact few patients have very large nodule or mass. This makes the pulmonary nodule and mass segmentation task resemble a long-tail problem rather than a mere large scale span problem. This leads to unsatisfactory results when segmenting large lesions that require more accurate delineation~\cite{yang2019probabilistic}.

Several studies have proposed solutions to tackle the large scale span challenges at both the input and feature level. For instance, some approaches adopt multi-scale inputs~\cite{chen2022multiresolution}, where the input images are resized to different resolution ratios. Some other methods leverage multi-scale feature maps to capture information from different scales, such as cross-scale feature fusion~\cite{tang2021lesion} or using multi-scale convolutional filters~\cite{chen2018encoder}. Furthermore, the attention mechanisms~\cite{vaswani2017attention} has also been utilized to emphasize the features that are more relevant for segmentation. 
Though these methods have achieved impressive performance, they still struggle to accurately segment the extremely imbalanced multi-scale lesions.

 Recently, some click-based lesion segmentation methods~\cite{tang2020one,tang2021lesion,tang2022accurate} introduce the click at the input or feature level and modify the network accordingly, resulting in higher accuracy results. Yet, the click input does not provide the scale information of lesions for the network. 

In this paper, we propose a scale-aware test-time click adaptation (SaTTCA) method, which simply utilizes easily obtainable lesion click (\ie, the center detected nodule) to adjust the parameters of the network normalization layers~\cite{DBLP:conf/iclr/WangSLOD21} during testing. Note that we do not need to exploit any data from the training set. Specifically, we expand the click into an ellipsoid mask, which supervises the test-time adaptation. This helps to improve the segmentation performance of large-scale nodules and masses. Additionally, we also propose a multi-scale input encoder to further address the problem of imbalanced lesion scales. 
Experimental results on two public datasets and one in-house dataset demonstrate that the proposed method outperforms existing methods with different backbones.

\section{Method}

\begin{figure}[tb]
	\includegraphics[width=\linewidth]{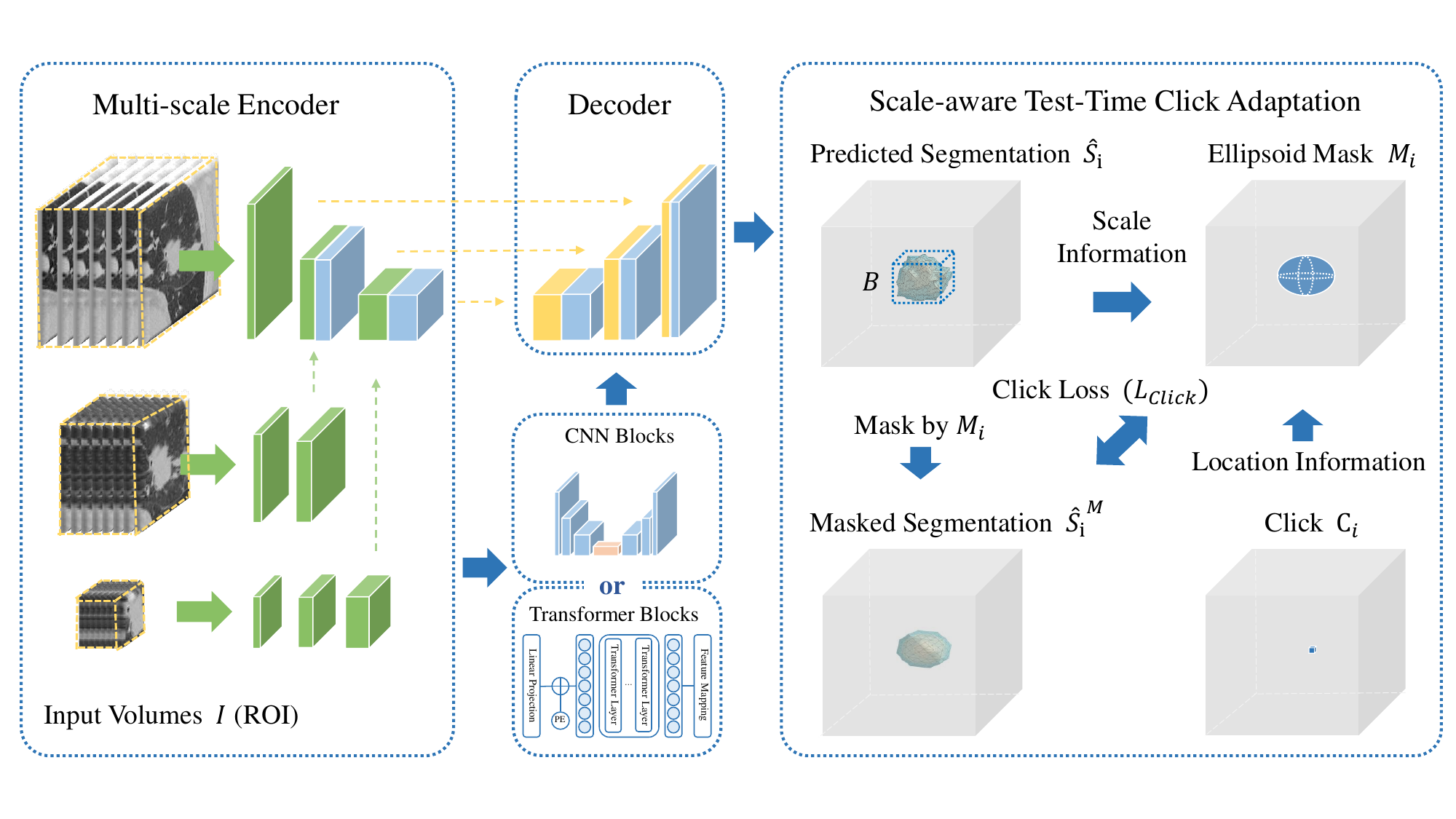}
	\caption{The pipeline of the proposed Scale-aware Test-time Click Adaptation (SaTTCA). We first get the predicted segmentation $\hat{S}_i$ and compute its minimum 3D bounding box $B$ from the trained model. Then we generate an ellipsoid mask $M_i$ (around the center $C_i$ of the detected nodule) whose size is proportional to the size of $B$ to supervise the parameter updating during test-time adaptation.
    Our SaTTCA method is applicable to backbones on CNN and Transformer. We also adopt a multi-scale input encoder to further improve the segmentation performance of nodules and masses with different scales.}
	\label{fig:method}
\end{figure}

\subsection{Restatement of image segmentation based on click}

For pulmonary nodule and mass segmentation, existing methods mostly rely on regions of interest (ROI) obtained by lesion detection networks. A set of 3D ROI inputs $I$ can be represented as $I \in \mathbb{R}^{D \times H \times W}$ with size $(D, H, W)$, along with its corresponding segmentation ground truth of nodules and masses represented by ${S}\in {(0,1)}^{D \times H \times W}$. Typically, a neural network with weighted parameters $\mathbf{\theta}$ is trained to predict the lesion area ${\hat{S}} = {\theta}({I})$, with the goal of minimizing the loss function $\mathcal{L}\:( S,\:\hat{S} )$. The stochastic gradient descent (SGD) and the automatic data acquisition module weight decay (AdamW) optimizers are usually used to optimize the weighted parameters.

 For each ROI input, the center point ${C}$ of the lesion, which is represented as $P_c = ( \tfrac{D}{2},\:\tfrac{H}{2},\:\tfrac{W}{2} )$ in Cartesian coordinate system, can be used as a reference point to assist the network in improving segmentation performance. This can be achieved either through an artificial or automatic approach, for instance, by adding click channels directly to the input or by adding a prior encoder to the network as demonstrated by the methods~\cite{tang2020one,tang2021lesion}. However, incorporating clicks in this way does not focus on addressing the extremely imbalanced lesion scales.

\subsection{Network architecture} 
The network structure of the proposed method, as shown in Fig.~\ref{fig:method}, is enhanced with a multi-scale (MS) input encoder to address the issue of multi-scale lesions. To achieve this, we employ a clipping strategy to adjust the proportion of foreground and background in the input image, producing a group of input images with dimensions of $64\times 96 \times96$, $32\times 48 \times48$, and $16\times 24 \times24$. These images are then passed through three convolution paths. The feature maps are concatenated as they are down-sampled to the same scale. The subsequent modules can be based on either CNN or transformer structures. The multi-scale input encoder allows the network to capture more scale information of the nodules and masses, thus mitigating the problem of large lesion scale span.

\subsection{Scale-aware test-time click adaptation} 
In clinical scenarios, the neural network for assisted diagnosis is generally a pre-training model. Due to differences in the statistical distribution of pulmonary nodule scale in image data from different medical centers, the segmentation results of some images, especially for large nodules, are worse than expected. For such scenarios, we propose the Scale-aware Test-time Click Adaptation method, which can improve the performance of segmentation results for large-scale nodules and masses by adjusting some of the network parameters during testing. The pipeline of the proposed method is shown in Fig.~\ref{fig:method}. First, we use the pre-trained network to pre-segment the input CT from the test set, getting ${\hat{S}_i} = \mathbf{\theta} \:(I_i)\:(i = 1, 2, \cdots, n)$ where n is the number of samples in the test set. Then we make a projection on the main connected region of ${\hat{S}_i}$ along three coordinate axes to obtain the size of the bounding box ${B_i} = (d, w, h)$ of the pre-segmentation result, and generate an ellipsoid ${M}_i$ with three axes length proportional to the corresponding side length of the bounding box $B_i$. More formally, the coordinates of any foreground voxel point $V : (x, y, z)$ in $M_i$ meets the following requirement:
\begin{equation}
\frac{(x - \tfrac{D}{2})^2}{\mathcal{R}\:(d)^2} + \frac{(y - \tfrac{H}{2})^2}{\mathcal{R}\:(h)^2}+ \frac{(z - \tfrac{W}{2})^2}{\mathcal{R}\:(w)^2} = 1, 
\end{equation}
where $\mathcal{R}$ represents the mapping function between the axis length of the ellipsoid and the side length of the bounding box $B_i$. Taking the x-axis as an example, $\mathcal{R}(d)$ is given by:
\begin{equation}
\mathcal{R}\:(d) = min\:(0.02\times d^2,\quad0.8 \times d).\label{con:mapping function}
\end{equation}

To account for the introduction of error information at some voxels during adaptive click adjustment, we develop a mapping function to generate $M_i$ adaptively based on the size of nodules and masses. If the nodule's length and diameter are less than 7 mm, ${M}_i$ degenerates into a voxel. When the predicted nodule size ranges from 7 mm to 40 mm, the axial length of $B_i$ and the side length of the bounding box follow a quadratic nonlinear relationship. If the predicted nodule size is greater than 40 mm, the axial length of $M_i$ has a linear relationship with the side length. To determine the super parametric values for the mapping function $\mathcal{R}$, we perform cross-validation on three datasets.

\subsection{Training objective of SaTTCA} 
We use the foreground range of adaptively adjusted ellipsoid $M_i$ to mask $\hat{S}_i$ to obtain a masked segmentation ${\hat{S}_i}^M$. Then we use ${M}_i$ to adjust the normalization layer parameters in the network during testing~\cite{DBLP:conf/iclr/WangSLOD21}. The test-time loss function $\mathcal{L}_{tt}$ is the weighted sum of the binary cross-entropy loss $\mathcal{L}_{BCE}$ and the Dice loss with sigmoid $\mathcal{L}_{Dice}$ of ${M}_i$ and ${\hat{S}_i}^M$, and the information entropy loss $\mathcal{L}_{ent}$ of ${\hat{S}_i}$. Formally, $\mathcal{L}_{tt}$ is given by:
\begin{equation}
\mathcal{L}_{tt} = \mathcal{L}_{BCE} + \sigma\mathcal{L}_{Dice}+\gamma\mathcal{L}_{ent},
\end{equation}

where $\sigma$ and $\gamma$ are hyper-parameters set to 0.5 and 1 in all experiments, respectively. 
The sum of the first two equations is referred to as click loss $\mathcal{L}_{Click}$.

\section{Experiments}
\subsection{Datasets and evaluation protocols}
We experiment on two public datasets and one in-house dataset. All three datasets are divided into training, validation, and test sets using a 7:1:2 ratio.
\subsubsection{LIDC~\cite{armato2011lung}:}
The LIDC dataset is a publicly available lung CT image database containing 1018 scans, developed by the Lung Image Database Consortium (LIDC). All pulmonary nodules and masses in the dataset have been annotated by multiple raters. To generate the ground truth for each nodule and mass, we combined the segmentation annotations from different raters. Overall, we selected a total of 1625 nodules and masses that were annotated by more than three raters from the LIDC dataset for the experiment.
\subsubsection{LNDb~\cite{pedrosa2019lndb}:}
The LNDb dataset published in 2019, comprises 294 CT scans collected between 2016 and 2018. Each CT scan in the dataset has been segmented by at least one radiologist. The nodules included in this dataset are larger than 3 mm. The mean scale of the lesion in LNDb dataset is the shortest among the three datasets. We adopt 1968 nodules and masses from the LNDb dataset.
\subsubsection{In-house data (ours):}
The in-house data (ours) contains 4055 CT scans and 6864 nodules and masses. Every CT scans are annotated with voxel-level nodule masks by radiologists. We exclude nodules and masses with diameters larger than 64 mm or smaller than 2 mm, as the diameter of the largest mass in the public dataset is no more than 64 mm.

\subsubsection{Evaluation Metrics:}
The performance of the nodule segmentation is evaluated by three metrics: volume-based Dice Similarity Coefficient (DSC), surface-based Normalized Surface Dice (NSD)~\cite{nikolov2018deep}, and recall rate, which calculates the shape similarity between predictions and ground truth.

\begin{table}[tb]
    \scriptsize
    \caption{{Performance of different backbones with or without the proposed SaTTCA and other click-based methods.} Experiments are conducted with various pulmonary nodule segmentation datasets  using 3D nnUNet~\cite{isensee2021nnu}, TransBTS~\cite{wang2021transbts} and nnUNet with multi-scale input encoder (MS) as the backbone. Comparative experiments are carried out with the click-based methods in~\cite{tang2020one,tang2021lesion} and simple test-time click adaptation on MS-UNet.  
    }\label{tab:metrics}
	\centering
    \begin{tabular*}{\hsize}{@{}@{\extracolsep{\fill}}l|ccc|ccc|ccc@{}}
 
    \toprule
    \multirow{2}{*}{Backbones}&\multicolumn{3}{c|}{LIDC}&\multicolumn{3}{c|}{LNDb}& \multicolumn{3}{c}{In-House}\\ 
    & DSC$\uparrow$ & NSD$\uparrow$ & Recall$\uparrow$ & DSC$\uparrow$ & NSD$\uparrow$ & Recall$\uparrow$ & DSC$\uparrow$ & NSD$\uparrow$ & Recall$\uparrow$\\ 

    \midrule    
    TransBTS~\cite{wang2021transbts} & 71.42 & 88.76 & 81.58 &	64.91 &	90.69 &	78.99 &	73.48 &	88.49 &	81.45\\
    TransBTS + SaTTCA & 72.08 &	89.55 &	82.46&65.72&91.89 &	$\mathbf{80.18}$&74.52&	89.53& 82.09\\ 
    \midrule
    nnUNet~\cite{isensee2021nnu} & 74.86 & 92.12 & 82.32 & 70.30 & 94.86 & 75.28 & 77.88 & 92.37 & 81.68\\
    nnUnet + SaTTCA & 75.62 & 93.05 & $\mathbf{83.61}$ & $\mathbf{71.46}$& $\mathbf{96.01}$& 78.68& 78.87 & 93.55& 83.53\\
    \midrule    
    nnUNet + MS  & 76.62 & 93.75 &	81.21 & 69.82 &	94.69 &	75.82 &	78.54 &	93.07 &	83.15\\ 
    nnUNet + MS +~\cite{tang2020one} & 76.96 &	92.39 &	78.53&	69.56&	93.76 &	76.11 &	77.39 &	92.03 &	79.14\\
    nnUNet + MS +~\cite{tang2021lesion} & 71.97 &91.78 & 73.90 & 67.62 & 91.02& 71.75 & 75.84 & 91.10 & 77.74\\
    nnUNet + MS + TTCA & 76.74& 93.85& 82.03& 69.81& 94.69 & 75.81 & 78.66& 93.19& 83.43\\
   nnUNet + MS + SaTTCA &$\mathbf{77.40}$&	$\mathbf{94.63}$ &	82.60&	70.62&	95.04 &	76.09&$\mathbf{79.62}$&$\mathbf{94.09}$ &	$\mathbf{85.04}$\\
    \bottomrule
    \end{tabular*}
\end{table}

\subsection{Implementation details}
The ROI of the lesion is a patch cropped around nodules or masses from the original CT scans with shape $64 \times 96 \times 96$. During pre-processing, Hounsfield Units (HU) values in all patches are first clipped to the range of $[-1350, 150]$. Min-max normalization is then applied, scaling HU values into the range of $[0, 1]$. All models are trained using AdamW~\cite{Loshchilov2019DecoupledWD} optimizer, cosine annealing learning rate schedule~\cite{Loshchilov2017SGDRSG} from $10^{-3}$ to $10^{-6}$ and batch size of 32. The training epoch is set to 200, and the test-time training epoch is set to 10. All experiments are conducted on 4 NVIDIA RTX 3090 GPUs with PyTorch 1.11.0~\cite{Paszke2019PyTorchAI}.

\begin{table}[tb]
    \scriptsize
    \caption{The performance variation (\%) of using Test-time Click Adaptation (TTCA) and Scale-aware TTCA (SaTTCA) on the nnUNet~\cite{isensee2021nnu} with multi-scale input encoder. 
    }\label{tab:metrics2}
	\centering
    \begin{tabular*}{\hsize}{@{}@{\extracolsep{\fill}}l|ccc|ccc|ccc@{}}
 
    \toprule
    \multirow{2}{*}{Scale}&\multicolumn{3}{c|}{LIDC}&\multicolumn{3}{c|}{LNDb}& \multicolumn{3}{c}{In-House}\\ 
    &$\Delta$ DSC &$\Delta$ NSD &$\Delta$ Recall &$\Delta$ DSC &$\Delta$ NSD &$\Delta$ Recall &$\Delta$ DSC &$\Delta$ NSD &$\Delta$ Recall\\ 
    \midrule
    \multicolumn{10}{l}{\textbf{\textit{w/ Test-time Click Adaptation}}}\\
    Micro&-0.034 &-0.072 &0.440 &-0.078 &-0.017 &-0.089 &-0.063 &-0.042 &0.334\\
    Small &0.042	&-0.087 &0.824	&0.038	&0.066	&0.218 &-0.016 &-0.046 &0.399\\
    Medium &0.147	&0.114	&0.795	&0.136	&0.195 &0.217 &0.040 &0.023 &0.431\\
    Mass &0.530	&0.465	&0.994	&-0.007 &-0.006	&-0.002 &0.107 &0.099 &0.378\\
    \midrule
    \multicolumn{10}{l}{\textbf{\textit{w/ Scale-aware Test-time Click Adaptation}}}\\
    Micro&0.593 &0.642 &1.256 &0.783 &0.277 &0.253 &0.900 &0.892 &0.518\\
    Small &0.831 &0.944 &1.547	&0.740	&0.296	&0.278	&1.109 &1.045 &0.881\\
    Medium &1.337 &1.892 &1.693	&0.930 &1.014 &0.964 &1.324 &1.219&1.646\\
    Mass &2.963 &3.182 &2.701 &2.590 &3.558 &3.093 &1.710 &1.656 &2.676\\
    \bottomrule
    \end{tabular*}
\end{table}

\subsection{Results}

We adopt nnUNet~\cite{isensee2021nnu} and TransBTS~\cite{wang2021transbts} as the backbone to evaluate the proposed method on pulmonary nodule and mass segmentation. nnUNet is a robust baseline with a complete CNN structure. Its adaptive framework makes it well-suited for pulmonary nodule segmentation. TransBTS is a 3D medical image segmentation network with a hybrid architecture of transformer and CNN. It incorporates long-range dependencies into the traditional CNN structure to achieve a larger receptive field. The experimental results presented in Tab.~\ref{tab:metrics}, consistently demonstrate that the CNN-based network can achieve better results in multi-scale pulmonary nodule and mass segmentation tasks across all three datasets. This is mainly due to the fact that large receptive fields may involve background features that are not conducive to segmentation inference for micro or small nodules. In datasets such as LIDC and In-House, where the number imbalance of multi-scale lesion phenomena is more notable, the multi-input method consistently outperforms the other two baselines. 

\begin{figure}[tb]
	\includegraphics[width=\linewidth]{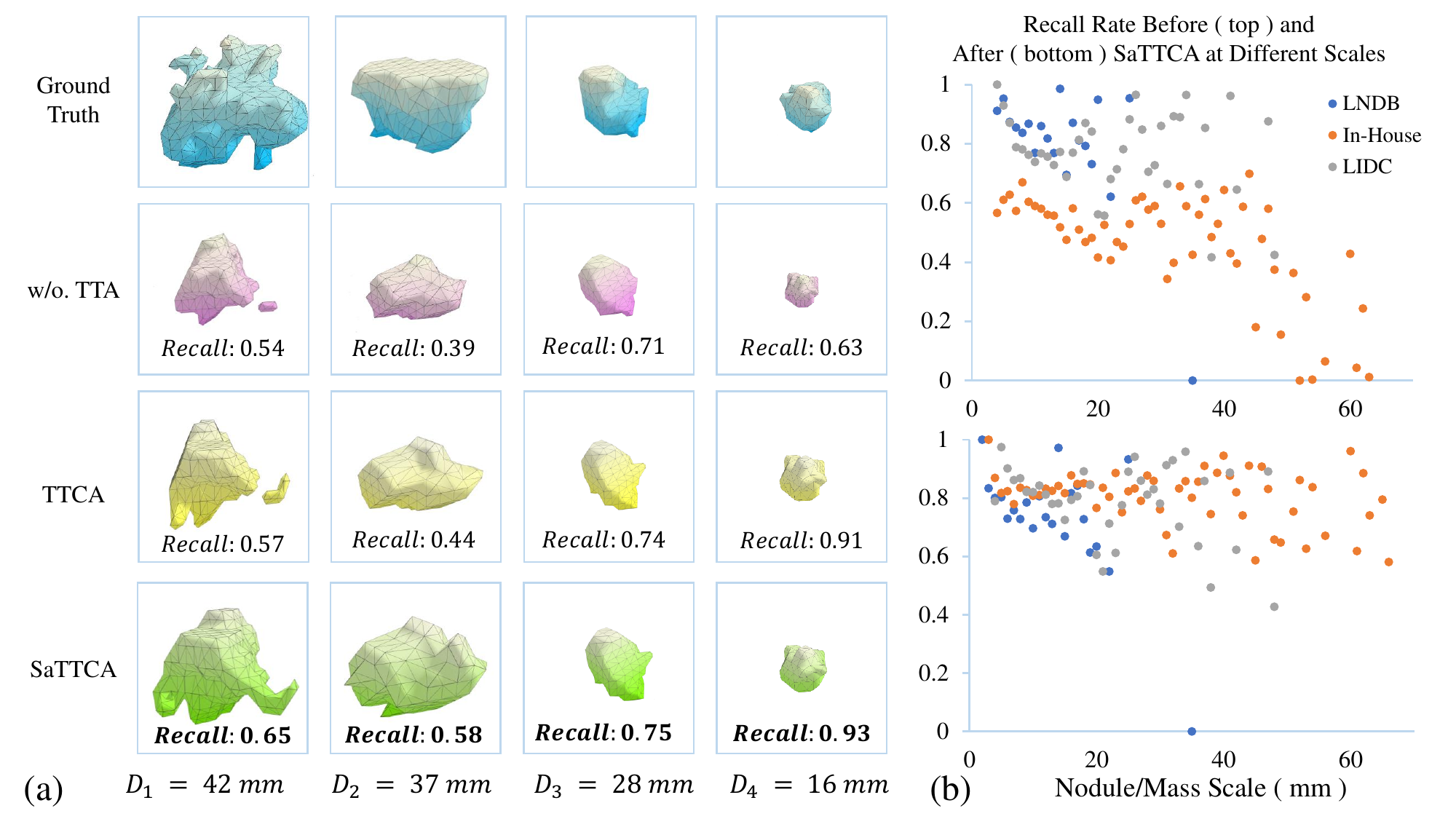}
	\caption{\textbf{(a)}: Visualization of some segmentation results predicted by the baseline without test-time adaptation (TTA), with test-time click adaptation (TTCA), and scale-aware test-time click adaptation (SaTTCA). The recall rates show that SaTTCA significantly improves the segmentation performance for large nodules and masses. \textbf{(b)}: The recall rate with respect to different scale of nodules/masses for the baseline method (top) and the proposed SaTTCA (bottom).}
	\label{fig:result}
\end{figure}

We also implemented comparative experiments with other click methods~\cite{tang2020one} and~\cite{tang2021lesion}. As depicted in Tab.~\ref{tab:metrics}, the experimental results show that using a point and a fixed range of gaussian intensity expansion in the case of large fluctuations in the size of the nodules does not take effect in improving the segmentation performance. The inferior segmentation results of~\cite{tang2021lesion} can be attributed to the fact that when it fuses features of different depths and scales, the number of channels in the feature map remains the same, and some of the up-sampling or down-sampling strides are too large, leading to redundancy in shallow features and a lack of deep features. Moreover, the SaTTCA improves the Dice coefficient and surface-based Normalized Surface Dice of segmentation results in both networks. In particular, as demonstrated in Fig.~\ref{fig:result}, the recall rate of large nodule segmentation is significantly improved.

We further analyze the performance of SaTTCA. Firstly, we present the quantitative comparison in  Tab.~\ref{tab:metrics2}, where we group the nodules and masses in each dataset at 10 mm intervals and calculate the average segmentation performance differences of the nodules in each scale group. The statistical results show that the proposed SaTTCA significantly improves the recall rate of the segmentation on large nodules and masses. As shown in Fig.~\ref{fig:result} (a), for nodules smaller than 20 mm, both TTCA and SaTTCA effectively increase the recall rate of predicted segmentation. For the medium nodule and mass, our SaTTCA proves to be more effective in improving segmentation performance. Fig.~\ref{fig:result} (b) shows the mean recall rate for lesions at every scale. The difference between the two scatter diagrams indicates that the proposed SaTTCA effectively alleviates the issue of extremely imbalanced lesion scales, and improves the segmentation performance for large lesions. In addition, for ten epochs of TTA, the inference time of each sample will increase approximately one second comparing with baseline.

\section{Conclusion}
This paper introduces a novel approach called the Scale-aware Test-time Click Adaptation for nodule and mass segmentation, which aims to address the issue of extremely imbalanced lesion scale and poor segmentation performance on large-scale nodules and masses. The network parameters are adapted at the instance level according to the scale-aware click during testing without altering the model architecture. This allows the network to achieve high recall for large-scale lesions. Then, a multi-scale input encoder is also proposed to enhance the segmentation performance of multi-scale nodules and masses. Extensive experiments on two public datasets and one in-house dataset demonstrate that though SATTCA increases inference time for each sample by about one second, it outperforms the corresponding baseline and click-based methods with different backbones.

\section{Acknowledgement}
This work was supported by the National Key Research and Development Program of China (2018AAA0100400), and in part by the National Natural Science Foundation of China (under Grants 62225113, 62222112 and 62176186).

\bibliographystyle{splncs04}
\bibliography{reference}

\begin{thebibliography}{10}
\providecommand{\url}[1]{\texttt{#1}}
\providecommand{\urlprefix}{URL }
\providecommand{\doi}[1]{https://doi.org/#1}

\bibitem{armato2011lung}
Armato~III, S.G., McLennan, G., Bidaut, L., McNitt-Gray, M.F., Meyer, C.R.,
  Reeves, A.P., Zhao, B., Aberle, D.R., Henschke, C.I., Hoffman, E.A., et~al.:
  The lung image database consortium (lidc) and image database resource
  initiative (idri): a completed reference database of lung nodules on ct
  scans. Medical physics  \textbf{38}(2),  915--931 (2011)

\bibitem{azad2022transdeeplab}
Azad, R., Heidari, M., Shariatnia, M., Aghdam, E.K., Karimijafarbigloo, S.,
  Adeli, E., Merhof, D.: Transdeeplab: Convolution-free transformer-based
  deeplab v3+ for medical image segmentation. In: Proc. of Intl. Conf. on
  Medical Image Computing and Computer Assisted Intervention. pp. 91--102
  (2022)

\bibitem{chen2018encoder}
Chen, L.C., Zhu, Y., Papandreou, G., Schroff, F., Adam, H.: Encoder-decoder
  with atrous separable convolution for semantic image segmentation. In: Proc.
  of European Conference on Computer Vision. pp. 801--818 (2018)

\bibitem{chen2022multiresolution}
Chen, S., Qiu, C., Yang, W., Zhang, Z.: Multiresolution aggregation transformer
  unet based on multiscale input and coordinate attention for medical image
  segmentation. Sensors  \textbf{22}(10), ~3820 (2022)

\bibitem{gould2013evaluation}
Gould, M.K., Donington, J., Lynch, W.R., Mazzone, P.J., Midthun, D.E., Naidich,
  D.P., Wiener, R.S.: Evaluation of individuals with pulmonary nodules: When is
  it lung cancer?: Diagnosis and management of lung cancer: American college of
  chest physicians evidence-based clinical practice guidelines. Chest
  \textbf{143}(5),  e93S--e120S (2013)

\bibitem{Heuvelmans2013OptimisationOV}
Heuvelmans, M., Oudkerk, M., de~Bock, G.H., de~Koning, H.J., Xie, X., van
  Ooijen, P.M.A., Greuter, M.J.W., de~Jong, P.A., Groen, H.J.M., Vliegenthart,
  R.: Optimisation of volume-doubling time cutoff for fast-growing lung nodules
  in ct lung cancer screening reduces false-positive referrals. European
  Radiology  \textbf{23},  1836--1845 (2013)

\bibitem{isensee2021nnu}
Isensee, F., Jaeger, P.F., Kohl, S.A., Petersen, J., Maier-Hein, K.H.: nnu-net:
  a self-configuring method for deep learning-based biomedical image
  segmentation. Nature methods  \textbf{18}(2),  203--211 (2021)

\bibitem{li2020learning}
Li, Y., Yang, J., Xu, Y., Xu, J., Ye, X., Tao, G., Xie, X., Liu, G.: Learning
  tumor growth via follow-up volume prediction for lung nodules. In: Proc. of
  Intl. Conf. on Medical Image Computing and Computer Assisted Intervention.
  pp. 508--517 (2020)

\bibitem{liu2022stbi}
Liu, K.: Stbi-yolo: A real-time object detection method for lung nodule
  recognition. IEEE Access  \textbf{10},  75385--75394 (2022)

\bibitem{Loshchilov2017SGDRSG}
Loshchilov, I., Hutter, F.: Sgdr: Stochastic gradient descent with warm
  restarts. Proc. of International Conference on Learning Representations
  (2017)

\bibitem{Loshchilov2019DecoupledWD}
Loshchilov, I., Hutter, F.: Decoupled weight decay regularization. In: Proc. of
  International Conference on Learning Representations (2019)

\bibitem{macmahon2017guidelines}
MacMahon, H., Naidich, D.P., Goo, J.M., Lee, K.S., Leung, A.N., Mayo, J.R.,
  Mehta, A.C., Ohno, Y., Powell, C.A., Prokop, M., et~al.: Guidelines for
  management of incidental pulmonary nodules detected on ct images: from the
  fleischner society 2017. Radiology  \textbf{284}(1),  228--243 (2017)

\bibitem{nikolov2018deep}
Nikolov, S., Blackwell, S., Zverovitch, A., Mendes, R., Livne, M., De~Fauw, J.,
  Patel, Y., Meyer, C., Askham, H., Romera-Paredes, B., et~al.: Deep learning
  to achieve clinically applicable segmentation of head and neck anatomy for
  radiotherapy. arXiv preprint arXiv:1809.04430  (2018)

\bibitem{oizumi2011anatomic}
Oizumi, H., Kanauchi, N., Kato, H., Endoh, M., Suzuki, J., Fukaya, K.,
  Sadahiro, M.: Anatomic thoracoscopic pulmonary segmentectomy under
  3-dimensional multidetector computed tomography simulation: a report of 52
  consecutive cases. The Journal of Thoracic and Cardiovascular Surgery
  \textbf{141}(3),  678--682 (2011)

\bibitem{Paszke2019PyTorchAI}
Paszke, A., Gross, S., Massa, F., Lerer, A., Bradbury, J., Chanan, G., Killeen,
  T., Lin, Z., Gimelshein, N., Antiga, L., Desmaison, A., K{\"o}pf, A., Yang,
  E., DeVito, Z., Raison, M., Tejani, A., Chilamkurthy, S., Steiner, B., Fang,
  L., Bai, J., Chintala, S.: Pytorch: An imperative style, high-performance
  deep learning library. In: Proc. of Advances in Neural Information Processing
  Systems (2019)

\bibitem{pedrosa2019lndb}
Pedrosa, J., Aresta, G., Ferreira, C., Rodrigues, M., Leit{\~a}o, P., Carvalho,
  A.S., Rebelo, J., Negr{\~a}o, E., Ramos, I., Cunha, A., et~al.: Lndb: a lung
  nodule database on computed tomography. arXiv preprint arXiv:1911.08434
  (2019)

\bibitem{schuchert2007anatomic}
Schuchert, M.J., Pettiford, B.L., Keeley, S., D’Amato, T.A., Kilic, A.,
  Close, J., Pennathur, A., Santos, R., Fernando, H.C., Landreneau, J.R.,
  et~al.: Anatomic segmentectomy in the treatment of stage i non-small cell
  lung cancer. The Annals of thoracic surgery  \textbf{84}(3),  926--933 (2007)

\bibitem{sung2021global}
Sung, H., Ferlay, J., Siegel, R.L., Laversanne, M., Soerjomataram, I., Jemal,
  A., Bray, F.: Global cancer statistics 2020: Globocan estimates of incidence
  and mortality worldwide for 36 cancers in 185 countries. CA: a cancer journal
  for clinicians  \textbf{71}(3),  209--249 (2021)

\bibitem{tang2021lesion}
Tang, Y., Yan, K., Cai, J., Huang, L., Xie, G., Xiao, J., Lu, J., Lin, G., Lu,
  L.: Lesion segmentation and recist diameter prediction via click-driven
  attention and dual-path connection. In: Proc. of Intl. Conf. on Medical Image
  Computing and Computer Assisted Intervention. pp. 341--351 (2021)

\bibitem{tang2020one}
Tang, Y., Yan, K., Xiao, J., Summers, R.M.: One click lesion recist measurement
  and segmentation on ct scans. In: Proc. of Intl. Conf. on Medical Image
  Computing and Computer Assisted Intervention. pp. 573--583 (2020)

\bibitem{tang2022accurate}
Tang, Y., Zhang, N., Wang, Y., He, S., Han, M., Xiao, J., Lin, R.S.: Accurate
  and robust lesion recist diameter prediction and segmentation with
  transformers. In: Proc. of Intl. Conf. on Medical Image Computing and
  Computer Assisted Intervention. pp. 535--544 (2022)

\bibitem{vachani2022probability}
Vachani, A., Zheng, C., Liu, I.L.A., Huang, B.Z., Osuji, T.A., Gould, M.K.: The
  probability of lung cancer in patients with incidentally detected pulmonary
  nodules: clinical characteristics and accuracy of prediction models. Chest
  \textbf{161}(2),  562--571 (2022)

\bibitem{vaswani2017attention}
Vaswani, A., Shazeer, N., Parmar, N., Uszkoreit, J., Jones, L., Gomez, A.N.,
  Kaiser, {\L}., Polosukhin, I.: Attention is all you need. Proc. of Advances
  in Neural Information Processing Systems  \textbf{30} (2017)

\bibitem{DBLP:conf/iclr/WangSLOD21}
Wang, D., Shelhamer, E., Liu, S., Olshausen, B.A., Darrell, T.: Tent: Fully
  test-time adaptation by entropy minimization. In: Proc. of International
  Conference on Learning Representations. OpenReview.net (2021)

\bibitem{wang2021transbts}
Wang, W., Chen, C., Ding, M., Yu, H., Zha, S., Li, J.: Transbts: Multimodal
  brain tumor segmentation using transformer. In: Proc. of Intl. Conf. on
  Medical Image Computing and Computer Assisted Intervention. pp. 109--119
  (2021)

\bibitem{yang2019probabilistic}
Yang, J., Fang, R., Ni, B., Li, Y., Xu, Y., Li, L.: Probabilistic radiomics:
  ambiguous diagnosis with controllable shape analysis. In: Proc. of Intl.
  Conf. on Medical Image Computing and Computer Assisted Intervention. pp.
  658--666 (2019)

\bibitem{yang2020hierarchical}
Yang, J., Gao, M., Kuang, K., Ni, B., She, Y., Xie, D., Chen, C.: Hierarchical
  classification of pulmonary lesions: a large-scale radio-pathomics study. In:
  Proc. of Intl. Conf. on Medical Image Computing and Computer Assisted
  Intervention. pp. 497--507 (2020)

\end{thebibliography}

\end{document}